\documentclass{ieeetj}
\JournalAcronym{OJSP}
\usepackage{cite}
\usepackage{placeins}
\newif\ifreferencesused
\referencesusedfalse
\let\originalcite\cite
\renewcommand{\cite}[2][]{%
    \global\referencesusedtrue
    \if\relax\detokenize{#1}\relax
        \originalcite{#2}%
    \else
        \originalcite[#1]{#2}%
    \fi
}
\usepackage{amsmath,amssymb,amsfonts,acronym}
\usepackage{algorithmic}
\usepackage{booktabs}
\usepackage{graphicx,color}
\graphicspath{{figures/}}
\usepackage{textcomp}
\usepackage{xcolor}
\usepackage{hyperref}
\hypersetup{hidelinks}
\usepackage{algorithm,algorithmic}
\makeatletter
\acrodef{LP}[MtReNIRS-LP]{Pretrained model with added Linear classification head as probe}
\acrodef{MLPprobe}[MtReNIRS-MLP]{Pretrained model with added MLP classification head as probe}
\acrodef{LoRA}[MtReNIRS-LoRA]{Pretrained model adapted with LoRA weights}
\acrodef{Full-FT}[MtReNIRS-Full-FT]{Full Finetune of pretrained model}
\acrodef{Scratch-FT}[Scratch-FT]{Train model from scratch without pretraining}
\acrodef{Stat-SVM}[Stat-SVM]{Statistical-feature Support Vector Machine classifier}
\acrodef{CNN}[CNN]{Convolutional Neural Network classifier}
\acrodef{LSTM}[LSTM]{Long Short-Term Memory classifier}

\makeatother
\def\BibTeX{{\rm B\kern-.05em{\sc i\kern-.025em b}\kern-.08em
    T\kern-.1667em\lower.7ex\hbox{E}\kern-.125emX}}
\AtBeginDocument{\definecolor{NavyBlue}{RGB}{0,86,125}}

\def\OJlogo{\vspace{-4pt}$<$Society logo(s) and publication title will appear here.$>$}
\def\seclogo{\vspace{10pt}$<$Society logo(s) and publication title will appear here.$>$}
\def\authorrefmark#1{\ensuremath{^{\textbf{#1}}}}

\begin{document}

\receiveddate{XX Month, XXXX}
\reviseddate{XX Month, XXXX}
\accepteddate{XX Month, XXXX}
\publisheddate{XX Month, XXXX}
\currentdate{XX Month, XXXX}
\doiinfo{XXXX.2022.1234567}

\markboth{}{Chae {et al.}}

\title{MtReNIRS: A Self-Supervised Foundation Model for Transferable Neurovascular Representations}
\author{Tony Chae\authorrefmark{1}, Kleanthis Avramidis\authorrefmark{2}, Dimitra Emmanouilidou\authorrefmark{3}, and Ivan J. Tashev\authorrefmark{3}}
\affil{Department of Electrical Engineering and Computer Sciences, University of California, Berkeley, Berkeley, CA, USA}
\affil{Signal Analysis and Interpretation Lab, University of Southern California, Los Angeles, CA, USA}
\affil{Microsoft Research, Redmond, WA, USA}
\corresp{Corresponding author: Tony Chae (email: tonychae01@berkeley.edu).}
\authornote{This work was done during an internship at Microsoft Research.}

\begin{abstract}
Functional near-infrared spectroscopy (fNIRS) offers a promising approach for monitoring the hemodynamic correlates of neural activity in naturalistic settings. However, existing modeling methods for fNIRS are largely confined to small-scale, single-session data with low sampling rates, driven by high heterogeneity in acquisition devices and optode montages. In addition, the complex spatiotemporal structure of fNIRS signals further restricts model scalability and transferability across settings. As a result, it remains unclear whether a single fNIRS encoder can learn representations that are usable across subjects, tasks, devices, and montages. Here, we show that a montage-aware masked autoencoder, pretrained on 12 public fNIRS datasets spanning 268 subjects and 567 recording hours, learns generalizable neurovascular representations that transfer across three external datasets. Our fNIRS foundation model, MtReNIRS, jointly encodes the slow hemodynamic nature of fNIRS, chromophore identity and cortical location by mapping heterogeneous montages onto a shared anatomical reference. MtReNIRS achieves cross-subject performance on par or better than fully supervised baselines and excels in few-shot within-subject calibration, where it outperforms prior state-of-the-art fNIRS models, while remaining robust to sparse montages. Our results indicate that montage-aware self-supervision yields physiologically grounded, transferable fNIRS representations, providing a scalable foundation for label-efficient brain-computer interfaces.

\end{abstract}


\maketitle

\begin{figure*}[!t]
    \centering
    \includegraphics[width=0.95\textwidth]{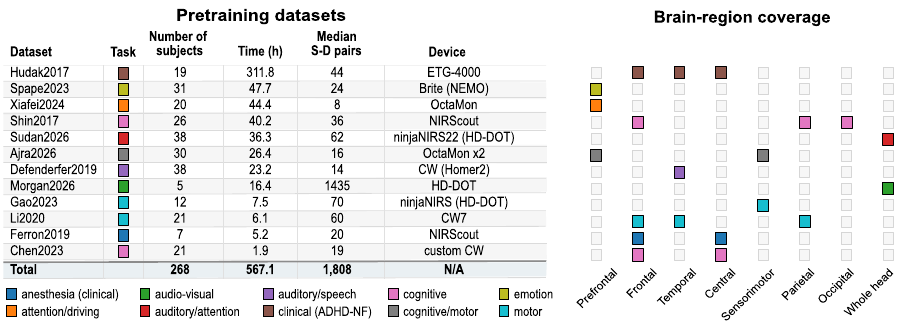}
    \caption{Heterogeneity of the 12 open-source fNIRS pretraining datasets: The table on the left reports  number of subjects, recording duration, median source--detector (S-D) pair count across processed runs, acquisition device, and task type; the matrix on the right portrays the brain-region coverage.}
    \label{fig:dataset-heterogeneity}
\end{figure*}

\section{Introduction}
Functional near-infrared spectroscopy (fNIRS) is a neuroimaging technology that measures cortical changes in oxygenated (HbO) and deoxygenated hemoglobin (HbR). Its non-invasive and portable sensors offer a promising way to expand brain-computer interfaces outside the lab and in everyday scenarios. Compared to electroencephalography (EEG), the fNIRS optical sensors can provide a complementary view of brain function with higher spatial resolution, while being more robust to myoelectric artifacts~\cite{bunterngchit2024towards,finnis2025exploring,jung2025efrm}. 

Collecting and annotating fNIRS data is costly and time-consuming, and usually requires domain expertise. Further, the scalability of computational modeling methods is often hindered by the necessary subject-specific calibration, which makes self-supervised foundation models particularly attractive. Prior work using large-scale pretraining on neural data has shown improved performance, generalization, and label efficiency, though existing pretrained models have predominantly targeted EEG~\cite{el2026reve,zhou2026csbrain,wang2025cbramod} and fMRI~\cite{wang2026towards,d2026foundation} modalities.

Comparable evidence for modeling fNIRS signals remains very limited. Fully supervised Transformer-based classifiers capture temporal dependencies within individual datasets~\cite{wang2022transformer,jing2023transformer}, while early fNIRS foundation-model studies have focused mainly on within-subject few-shot classification~\cite{jung2025fnirs,jung2025efrm}. It therefore remains unclear whether a single pretrained fNIRS encoder can learn representations that transfer across subjects, tasks, acquisition devices, and sensor montages.  Building a faithful fNIRS foundation model is challenging because fNIRS data are highly heterogeneous across studies \cite{srinivasan2024subject,feng2025heterogeneous}. Acquisition devices, sampling rates, signal representations, source--detector (S--D) geometries, and cortical coverage vary substantially, while individual datasets remain relatively small. A model trained on a single study may therefore learn study-specific acquisition characteristics rather than reusable neurovascular structure.

Scaling fNIRS requires more than pooling unlabeled recordings. Spatial channels across different montages must be aligned to a common anatomical reference so that each fNIRS signal can be consistently related to the underlying cortical activity. The pretraining objective should also reflect fNIRS physiology. HbO/HbR signals are slow, delayed, and highly autocorrelated, which differ fundamentally from the fast electrophysiology that many brain-signal encoders were designed for~\cite{wang2023rethinking}. To our knowledge, prior fNIRS pretraining models have not jointly combined montage-aware scale, physiology-matched mask reconstruction, and cross-subject evaluation. This motivates our central question: \textit{Can we learn transferable neurovascular representations at scale?}

We introduce \textbf{MtReNIRS} (\emph{\textbf{M}ontage-aligned
\textbf{t}ransferable \textbf{Re}presentations for f\textbf{NIRS}}), a foundation model that learns neurovascular representations with a montage-aware masked autoencoder. To promote generalizable representations, we train on 12 public datasets selected to maximize task, device, and subject diversity, spanning 268 subjects, 567.1 recording hours, and broad cortical coverage (Fig.~\ref{fig:dataset-heterogeneity}). We map heterogeneous S-D midpoint geometries to a shared standard 10--05 reference \cite{oostenveld2001five} and combine each 10-s HbO/HbR segment with its canonical coordinate and chromophore identity. The model then reconstructs randomly masked windows without knowledge of task labels. Transferability is evaluated on three unseen datasets of cognitive, auditory, and motor tasks.
Our main contributions are: 
\begin{itemize}
    \item[(1)] We develop a montage harmonization framework that maps heterogeneous fNIRS recordings to a shared anatomical representation, enabling joint pretraining across datasets with different sensor layouts. The framework supports 3-D coordinate alignment, anatomically anchored 2-D placement, and direct label assignment depending on the available montage information. 
    \item[(2)] We introduce a compact encoder of 1.23 M parameters that jointly represents fNIRS signal dynamics, anatomical location, and chromophore identity to learn physiologically grounded and transferable representations.
    \item[(3)] We systematically evaluate knowledge transfer on three unseen datasets through cross-subject leave-one-subject-out (LOSO) classification and within-subject calibration using only 2--8 labels per class. MtReNIRS outperformed most existing fNIRS pretraining methods~\cite{jung2025fnirs,jiang2021self,thapa2024sleepfm,jung2025efrm} and fully supervised approaches~\cite{wang2022transformer}.
\end{itemize}
Finally, we additionally examine the effects of token duration, spatial correspondence, pretraining duration, and robustness to reduced-channel settings.

\section{Related Work}
Supervised fNIRS models range from classifiers built on summary statistics to convolutional, recurrent, and attention-based networks~\cite{bunterngchit2024towards,jing2023transformer}. Wang et al. \cite{wang2022transformer} introduced fNIRS-T for supervised
multichannel hemodynamic decoding, while Lee et al. \cite{lee2024transformer}
extended this framework with optode-layout-aware spatial embeddings.
However, both operate within fixed, dataset-specific input spaces, limiting
transfer across devices and montages. Our approach preserves S-D pair identity
while accommodating variable sensor counts and arrangements.

BERT-style masked prediction in language and ViT-based masked autoencoding in computer vision established the objective of reconstructing hidden inputs from
visible context~\cite{devlin2019bert,he2022masked}. This
principle has since yielded useful representations for sleep, multimodal
physiology, and fNIRS~\cite{jiang2021self,thapa2024sleepfm,jung2025fnirs,jung2025efrm}. Related EEG-BCI studies have also explored cross-subject decoding and generalization across datasets \cite{monesi2024auditory,thornton2024decoding}.
However, existing fNIRS pretraining remains limited in corpus scale and focuses
primarily on within-subject few-shot classification. Generalization across
unseen subjects, tasks, devices, and reduced montages remains underexplored. More importantly, current pretraining models remain insufficiently grounded in delayed hemodynamics, coupled HbO/HbR responses, and montage anatomy \cite{wang2023rethinking}.

\begin{figure*}[!t]
    \centering
    \includegraphics[width=0.87\textwidth]{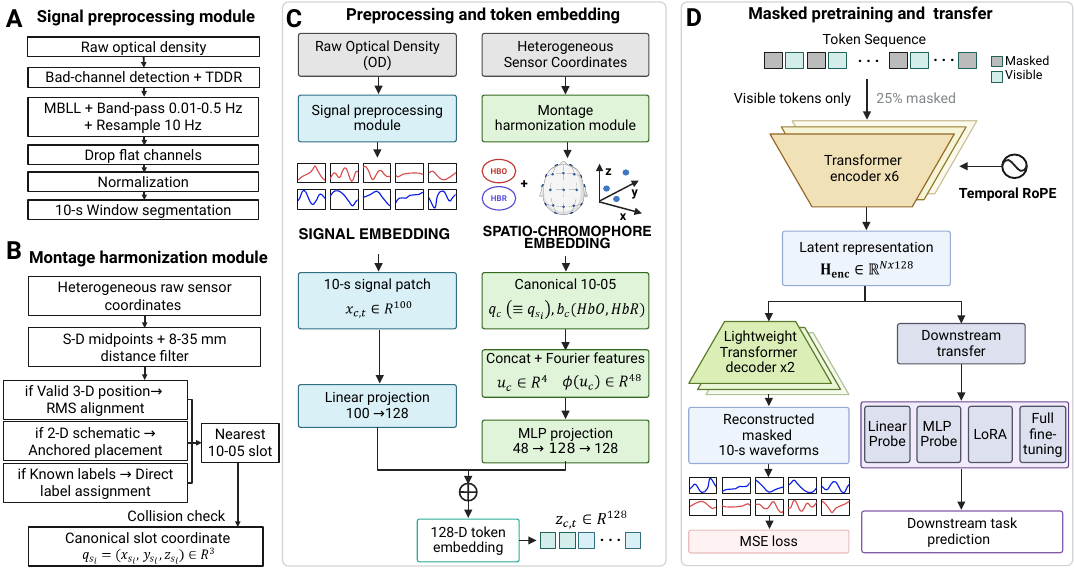}
    \caption{Overview of (A) fNIRS pre-processing pipeline, (B) harmonization pipeline of heterogeneous montages, (C) deriving signal and spatio-chromophore token embedding, and (D) masked pretraining and downstream transfer operations of the foundation model.}
    \label{fig:model-architecture}
\end{figure*}

\section{Datasets and Methods}

\subsection{Pretraining Datasets}
We assembled 12 public fNIRS datasets that varied in device, task, cortical
coverage, sampling rate, measurement density, and signal representation~\cite{ds007554:1.0.0,chen2023open,Defenderfer_Buss_2019,ds004541:1.0.0,ds007420:1.0.2,DANS-XRK-HNHB_2017,ds005935:1.0.0,ds007473:1.0.0,shin2018simultaneous,spape2023nemo,ds007738:1.0.0,ds004973:1.0.1}, as described in 
Fig.~\ref{fig:dataset-heterogeneity}. The datasets include tasks with
cognitive, motor, auditory, affective, clinical, and naturalistic recordings
from prefrontal, frontal, temporal, central, sensorimotor, parietal, occipital,
and whole-head coverage. The overall pretraining dataset contained 268
subject entries, 1,389 runs, and over 567 recording hours total. After quality control, two runs were excluded for poor signal quality. We segmented these runs into
203,420 non-overlapping 10-s windows. We treated each channel as independent inputs regardless of subject identity. A subject-disjoint split within each dataset assigned approximately 80\% of entries to training and 20\% to validation, which yielded 213 training and 53 validation entries in total.

Separately, we reserved three additional datasets exclusively for downstream evaluation. None of the downstream subjects or recordings were included in either
training or validation sets used for pretraining. These datasets were chosen to cover different subject counts and task types, while also considering datasets used in prior literature benchmarks \cite{bauernfeind2011single}, so we could test how well the learned representations can transfer across different settings. \textbf{Gunther2014}: 8--subject corpus containing recordings of mental arithmetic tasks and resting states~\cite{bauernfeind2011single}; \textbf{Shader2021}: 8--subject corpus containing auditory, visual, and
control speech tasks \cite{shader2021use}; and \textbf{SujinBak2019}: 30--subject corpus with left-hand, right-hand, and foot motor trials
~\cite{bak2019open}. Exact event extraction yielded 348, 368, and 2,250
10-s samples, respectively. 

We used 10-s segments throughout the study to capture the multi-second rise and peak of the local hemodynamic response while preserving finer temporal structure. We limited individual windows to 10-s because longer windows could extend beyond the labeled event boundaries in the downstream datasets and potentially mix adjacent conditions.

\subsection{Signal Pre-processing and Montage Harmonization}

As shown in Fig. \ref{fig:model-architecture}A, we applied a dataset-adapted preprocessing pipeline to recordings, provided either as raw intensity, optical density, or HbO/HbR concentrations. We retained only valid S-D pairs that were separated by 8--35~mm, when this information was available in the raw files. We then converted raw continuous-wave intensity to optical density, marked
wavelength channels with a scalp-coupling index below 0.5, applied temporal
derivative distribution repair, and used the modified Beer--Lambert law with a partial pathlength factor of 6.0. Runs were rejected only when no channel reached scalp-coupling index of 0.5. 

For recordings provided as raw intensity, we applied the full preprocessing pipeline; for those only provided with processed HbO/HbR concentrations, we skipped the preceding conversion steps. We then band-pass filtered each run from 0.01 to 0.5~Hz, resampled to 10~Hz, and removed flat or abnormally low-variance chromophore channels. This branching preserved each dataset's available signal information while producing a common HbO/HbR representation.
We z-scored each retained chromophore channel over its complete run using the
population mean and standard deviation, then clipped model inputs to $[-6,6]$.
Pretraining windows did not overlap, and downstream examples used exactly
$[\text{onset},\text{onset}+10~\mathrm{s})$ without a hemodynamic-response shift.

\begin{figure}[!t]
    \centering
    \includegraphics[width=0.90\columnwidth]{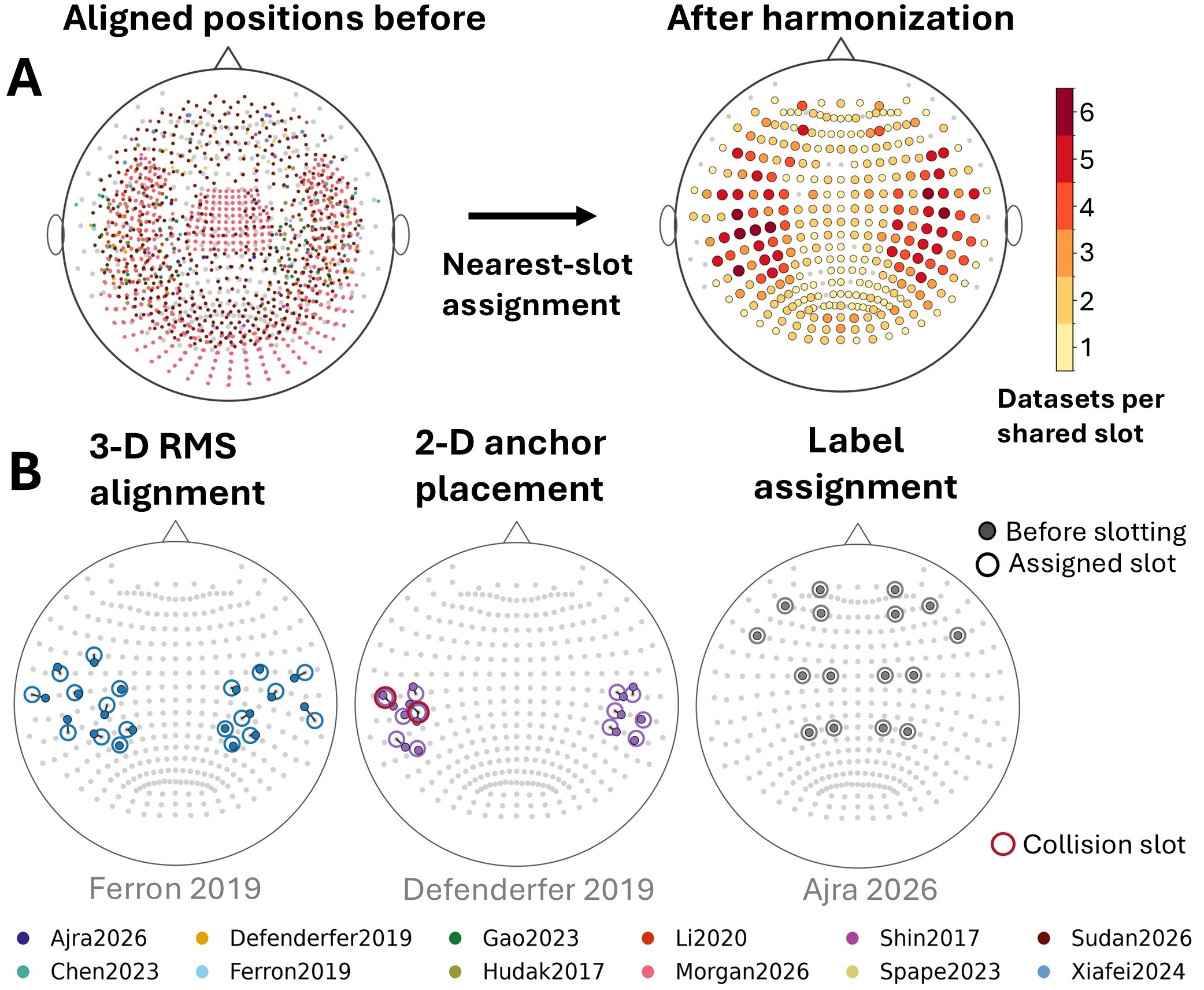}
    \caption{Montage-harmonization results. (A) Aligned source--detector midpoint positions before slotting and shared canonical 10--05 slots after nearest-slot assignment; color denotes the number of pretraining datasets represented at each slot. (B) Representative 3-D RMS alignment (Ferron 2019), 2-D anchored placement (Defenderfer 2019), and direct label assignment (Ajra 2026); red rings denote slot collisions.}
    \label{fig:montage-harmonization}
\end{figure}

Montage harmonization mapped each physical S-D pair to a shared canonical
10--05 slot (Fig. \ref{fig:model-architecture}B; Fig. \ref{fig:montage-harmonization}). We selected this denser
reference over standard 10--20 to reduce midpoint slot collisions while
retaining a scalp-coordinate vocabulary shared with EEG, which can support future EEG--fNIRS multimodal systems~\cite{oostenveld2001five}. For valid 3-D geometries provided by the dataset, we
centered the raw S-D midpoint coordinates $\mathbf p_i$ and matched their root
mean square (RMS) radius to the standard 10--05 reference (3-D RMS alignment; Fig. \ref{fig:montage-harmonization}B):
\begin{equation}
\tilde{\mathbf p}_i=(\mathbf p_i-\bar{\mathbf p})
\frac{s_q}{s_p}+\bar{\mathbf q},\qquad
s_p^2=\frac{1}{N}\sum_i\lVert\mathbf p_i-\bar{\mathbf p}\rVert_2^2.
\label{eq:rms-alignment}
\end{equation}
where $N$ is the number of valid S-D midpoints, $\bar{\mathbf p}$ and $\bar{\mathbf q}$ are the centers of the raw and canonical coordinates, respectively, and $s_p$ and $s_q$ denote the RMS distances from their respective centers.
We assigned each aligned midpoint to its nearest canonical slot. For datasets where only 2-D layout schematics were available, we placed the probe on an anatomically supported local tangent plane before nearest-slot assignment (2-D anchor placement; Fig. \ref{fig:montage-harmonization}B). For datasets with known placements but
missing or degenerate geometry, we assigned canonical labels (Label assignment; Fig. \ref{fig:montage-harmonization}B). For all positions created via the above procedures, we kept
the distinct S-D pairs even when they mapped to the same slot and recorded these events as slot collisions.

Our model used the fixed coordinate $\mathbf q_{s_i}$ of assigned slot $s_i$,
not the intermediate aligned coordinate. We normalized all available canonical
coordinates with one center and RMS scale computed from the pretraining corpus.
All 1,387 runs from the 12 pretraining datasets used these canonical spatial
inputs. We applied the same canonical-slot assignment and coordinate
normalization to every downstream input.

\subsection{Model and Self-Supervised Pretraining}
We represented each token by a chromophore channel $c$ and temporal window $t$, where $c=(i,b_c)$ identifies an S-D pair $i$ and chromophore type $b_c\in{0,1}$, with 0 and 1 denoting HbO and HbR, respectively.

We used 10-s (100-sample) windows to capture the multi-second rise and peak of a local hemodynamic response while retaining the temporal sequence of six 10-s windows in each 60-s crop, allowing the model to capture longer-range temporal relationships across windows through attention, consistent with evidence that temporal context and response timing matter for fNIRS decoding~\cite{wang2022transformer,wang2023rethinking}.
Each 10-s signal segment $\mathbf{x}_{c,t}\in\mathbb{R}^{100}$ was projected by a linear layer to a 128-dimensional signal embedding. For channel $c=(i,b_c)$, the canonical spatial coordinate is inherited from
its S-D pair, $\mathbf{q}_c=\mathbf{q}_{s_i}\in\mathbb{R}^3$.
We concatenated this coordinate with the binary chromophore indicator $b_c$
to form
$\mathbf{u}_c=[\mathbf{q}_c^\top,b_c]^\top\in\mathbb{R}^4$. We encoded the resulting $\mathbf{u}_c$ with six bands of Fourier
features to better represent this low-dimensional spatio-chromophore domain
~\cite{tancik2020fourier}:
\begingroup
\setlength{\abovedisplayskip}{4pt}
\setlength{\belowdisplayskip}{4pt}
\setlength{\abovedisplayshortskip}{4pt}
\setlength{\belowdisplayshortskip}{4pt}
\begin{equation}
\phi(\mathbf u_c)=
\left[
\sin(2^k\pi u_{c,j}),
\cos(2^k\pi u_{c,j})
\right]_{
\substack{j=1,\ldots,4\\k=0,\ldots,5}}
\in\mathbb R^{48}.
\label{eq:fourier-features}
\end{equation}
\endgroup
A $48\!\rightarrow\!128\!\rightarrow\!128$ multilayer perceptron (MLP) with
GELU activation projected these features. This embedding lets the encoder
distinguish region-specific activity while modeling coupled, often opposing
HbO/HbR dynamics. We summed the signal and spatio-chromophore embeddings to
form token $\mathbf{z}_{c,t}\in\mathbb{R}^{128}$, as shown in Fig.~\ref{fig:model-architecture}.

The encoder contained six pre-layer-normalized Transformer blocks with a
128-dimensional state, four attention heads, a 512-dimensional MLP, and 0.1
dropout. Per-head layer normalization stabilized queries and keys. To preserve
cross-window order without conflating time with spatial identity, rotary
position encoding (RoPE)\cite{su2024roformer} with base 500 encoded the temporal-window index. All
S-D pairs from the same window shared that index; spatial identity entered only
through the canonical coordinate embedding. The inference encoder contained
1.23 million parameters.

We trained the encoder as a visible-token masked autoencoder. For each 60-s
crop, we randomly masked 25\% of valid tokens and passed only visible tokens to
the encoder. A two-block lightweight Transformer decoder reconstructed each
masked 10-s waveform from the encoded visible tokens and learned mask tokens.
We minimized mean squared error only over masked real tokens,
\begingroup
\setlength{\abovedisplayskip}{4pt}
\setlength{\belowdisplayskip}{4pt}
\setlength{\abovedisplayshortskip}{4pt}
\setlength{\belowdisplayshortskip}{4pt}
\begin{equation}
\mathcal L_{\mathrm{MAE}}=
\frac{1}{|\mathcal M|\,100}
\sum_{n\in\mathcal M}\lVert\hat{\mathbf x}_n-\mathbf x_n\rVert_2^2.
\label{eq:mae-loss}
\end{equation}
\endgroup

For each training sample, we selected a dataset and a run uniformly, then drew a random 60-s crop containing six 10-s windows per chromophore channel. For recordings with more than 256 available chromophore channels, we randomly sampled 256 channels; recordings with 256 or fewer used all available channels. Each selected chromophore channel contributed six tokens, one per 10-s window, yielding a maximum of $(256 \times 6=1{,}536)$ tokens per crop. This sampling scheme gave each dataset equal probability of being selected, regardless of its number of subjects or recording hours. Validation used one fixed 60-s crop per run.

During pretraining, we trained the model for 100 epochs with AdamW, with a batch size of 128, a weight decay of 0.01, and a peak learning rate of $2\times10^{-4}$ with 10\% linear warm-up followed by cosine decay. We selected the epoch-95 checkpoint, which achieved the lowest self-supervised validation loss, without inspecting downstream labels. Each training run used one 32-GB NVIDIA Tesla V100 GPU; four GPUs executed separate
configurations in parallel.

\subsection{Downstream Evaluation Protocols}

Our evaluation comprised four adaptation strategies for pretrained MtReNIRS and four fully supervised baselines (no pretraining). \acs{LP} was used as the default setting when comparing with prior work, unless otherwise stated.

The adaptation strategies were: (a) {\textbf{\acs{LP}}}: the pretrained encoder was kept frozen, and only a linear classification head was trained; (b) {\textbf{\acs{MLPprobe}}}: the pretrained encoder was kept frozen, and an MLP
classification head with one 64-unit hidden layer and GELU activation was
trained; (c) {\textbf{\acs{LoRA}}}: low-rank adaptation of the pretrained encoder using rank-8
updates with $\alpha=16$ in the attention and feed-forward linear layers; (d) {\textbf{\acs{Full-FT}}}: the pretrained encoder and classification head were
jointly fine-tuned. The non-pretrained baselines were: (e) {\textbf{\acs{Scratch-FT}}}: the same MtReNIRS architecture was trained from random
initialization without pretraining; (f) {\textbf{\acs{CNN}}}: 1-D, two-layers, with 64 filters ($K=5$), global average pooling, and a linear head, trained on preprocessed HbO/HbR signals ; (g) {\textbf{\acs{LSTM}}}: bidirectional one-layer with 64 hidden units per direction, temporal mean pooling, and a linear head, trained on preprocessed HbO/HbR signals;
(h) {\textbf{\acs{Stat-SVM}}}: an RBF SVM ($C=1$) using eight temporal statistics per chromophore channel as input features. This model serves as a strong statistical
feature-engineering baseline for fNIRS with a nonlinear decision function \cite{heremans2025functional};

\textit{Benchmark 1: cross-subject transfer.} We held out one subject for testing
in each LOSO fold. Of the remaining non-test subjects, approximately 15\% were used for subject-level inner validation. Models were trained on the remaining subjects for 50 epochs with AdamW, an initial learning rate of $10^{-4}$ with cosine annealing, a weight decay of $10^{-4}$, and a batch size of 32. We evaluated the
state with the highest validation macro-F1 once on the held-out subject. The
fixed-hyperparameter Stat-SVM used all non-test subjects because it required no
epoch selection. We used the intersection of chromophore-channel names across
runs and preserved a common order within each dataset.

\textit{Benchmark 2: within-subject few-shot calibration.} For each subject, we randomly
sampled $N\in\{2,4,6,8\}$ examples without replacement from each class, repeated 50 times, and we tested on that subject's remaining examples. We used no other subjects, validation set, or early
stopping. \acs{LP} cached the frozen features and optimized a new linear head for 200 full-batch updates, following the learning rate and optimization of prior fNIRS few-shot studies to ensure a controlled comparison under matched training conditions. \acs{Scratch-FT} optimized a random backbone and head for 50
updates. \acs{CNN} and \acs{LSTM} used 200 updates, and \acs{Stat-SVM} was fitted using the same sampled examples.

To ensure comparable representations when token duration varied (2, 5, and 10-s), we temporally averaged encoder outputs before classification so that the feature dimensionality remained fixed. For the primary 10-s model, each trial contained a single window, making temporal-mean and flatten pooling equivalent.

\subsection{Metrics and Statistical Analysis}
We used macro-F1 as the primary metric and treated each subject as the
independent unit for statistical analysis. To help readers compare our findings with prior work, Table \ref{tab:gunther-prior-work} additionally reports
classification accuracy. For LOSO, we calculated each held-out
subject's score and reported the mean and population standard deviation across
subjects. For few-shot evaluation, we first averaged the 50 seed-level scores
within each subject and then summarized those subject means. Thus, repeated
few-shot draws from the same subject were not treated as independent samples.

For paired model comparisons, we calculated the macro-F1 difference between
the two models separately for each subject. We formed 95\% percentile
confidence intervals for the mean difference by resampling subjects with
replacement 20,000 times, and applied two-sided Wilcoxon signed-rank tests to
the subject-level paired differences. Reported $p$-values were not adjusted
for multiple comparisons.

\section{Experimental Results}

\subsection{Benchmark 1: Cross-Subject Transfer (LOSO)}

\begin{table}[!t]
    
    \centering
    \caption{Benchmark 1 cross-subject LOSO using the 10-s configuration. Macro-F1 (\%) is reported as the mean $\pm$ population standard deviation across held-out subjects. The fully supervised baselines use no pretraining. Boldface identifies the highest point estimate in each dataset; it does not denote statistical significance.}
    \label{tab:benchmark1-performance}
    \footnotesize
    \setlength{\tabcolsep}{2pt}
    \renewcommand{\arraystretch}{0.92}
    \begin{tabular*}{\columnwidth}{@{\extracolsep{\fill}}lccc@{}}
        \toprule
        \textbf{Method} & \shortstack{\textbf{Gunther 2014} \\ ($n=8$)} & \shortstack{\textbf{Shader 2021} \\ ($n=8$)} & \shortstack{\textbf{SujinBak 2019} \\ ($n=30$)} \\
        \midrule
        \multicolumn{4}{l}{\textbf{Fully supervised baselines}} \\
        \quad Scratch-FT    & 82.3 $\pm$ 7.7 & 40.2 $\pm$ 6.2 & 66.7 $\pm$ 16.2 \\
        \quad Stat-SVM      & 83.8 $\pm$ 7.4 & 37.2 $\pm$ 7.9 & 66.3 $\pm$ 18.2 \\
        \quad CNN           & 74.4 $\pm$ 9.6 & 34.4 $\pm$ 6.4 & 56.6 $\pm$ 15.9 \\
        \quad LSTM          & 64.3 $\pm$ 14.3 & 35.5 $\pm$ 8.1 & 59.8 $\pm$ 16.1 \\
        \addlinespace
        \multicolumn{4}{l}{\textbf{MtReNIRS pretrained}} \\
        \quad MtReNIRS-MLP     & 86.5 $\pm$ 7.7 & 44.7 $\pm$ 10.3 & 67.2 $\pm$ 18.6 \\
        \quad MtReNIRS-LoRA          & 86.2 $\pm$ 5.2 & 43.7 $\pm$ 11.6 & \textbf{70.3 $\pm$ 18.6} \\
        \quad MtReNIRS-Full-FT      & 83.8 $\pm$ 7.8 & \textbf{47.3 $\pm$ 5.6} & 70.2 $\pm$ 19.1 \\
        \quad MtReNIRS-LP            & \textbf{87.3 $\pm$ 7.8} & 44.5 $\pm$ 9.4 & 67.9 $\pm$ 17.7 \\
        \bottomrule
    \end{tabular*}
\end{table}

Across all three downstream datasets, Table~\ref{tab:benchmark1-performance}, all MtReNIRS adaptation strategies matched or exceeded the strong fully supervised baseline. Differences among adaptation strategies were relatively modest, although the leading point estimate varied by dataset: \acs{LP} on Gunther, \acs{Full-FT} on Shader, and \acs{LoRA} on SujinBak. Relative to \acs{Scratch-FT}, \acs{LP} gained 5.0 percentage points on Gunther (95\% CI: 0.2--10.0), 4.3 on Shader ($-2.6$--11.9), and 1.2 on SujinBak ($-0.8$--3.5), but none of the paired tests reached significance ($p=0.109$, 0.383, and 0.416). \acs{LP} also did not significantly exceed Stat-SVM ($p=0.266$, 0.055, and 0.237). It exceeded CNN and LSTM on Gunther (both $p=0.008$) and SujinBak (both $p<0.001$), and exceeded LSTM on Shader ($p=0.039$). \acs{LoRA} exceeded \acs{LP} on SujinBak by 2.4 points ($p=0.015$). Thus, pretraining supported competitive cross-subject transfer, while no single adaptation strategy was uniformly superior across datasets.

\subsection{Benchmark 2: Within-Subject Few-Shot Calibration}

\begin{table}[!t]
    \caption{Benchmark 2 within-subject few-shot. Macro-F1 (\%), reported as the mean $\pm$ population standard deviation in percentage points across subjects after averaging over 50 repetitions within each subject.}
    \label{tab:benchmark2}
    \centering
    \footnotesize
    \setlength{\tabcolsep}{2pt}
    \renewcommand{\arraystretch}{0.92}
    \begin{tabular*}{\columnwidth}{@{\extracolsep{\fill}}lcccc@{}}
        \toprule
        \textbf{Method} & \textbf{2-shot} & \textbf{4-shot} & \textbf{6-shot} & \textbf{8-shot} \\
        \midrule
        \multicolumn{5}{l}{\textbf{Gunther 2014} ($n=8$)} \\
        \quad Scratch-FT & 70.9 $\pm$ 12.6 & 76.7 $\pm$ 12.7 & 79.9 $\pm$ 11.2 & 82.8 $\pm$ 9.7 \\
        \quad Stat-SVM   & 42.7 $\pm$ 3.8 & 65.6 $\pm$ 9.9 & 72.6 $\pm$ 11.0 & 76.5 $\pm$ 11.0 \\
        \quad CNN        & 62.1 $\pm$ 10.6 & 68.9 $\pm$ 12.4 & 72.0 $\pm$ 13.1 & 74.8 $\pm$ 14.3 \\
        \quad LSTM       & 65.4 $\pm$ 10.4 & 69.6 $\pm$ 12.5 & 71.7 $\pm$ 12.4 & 74.1 $\pm$ 12.8 \\
        \quad MtReNIRS-LP       & \textbf{77.8 $\pm$ 11.6} & \textbf{84.4 $\pm$ 9.5} & \textbf{87.5 $\pm$ 8.0} & \textbf{89.6 $\pm$ 7.0} \\
        \addlinespace
        \multicolumn{5}{l}{\textbf{Shader 2021} ($n=8$)} \\
        \quad Scratch-FT & 34.8 $\pm$ 4.0 & 38.3 $\pm$ 5.9 & 39.7 $\pm$ 8.5 & 39.2 $\pm$ 9.6 \\
        \quad Stat-SVM   & 31.9 $\pm$ 3.8 & 36.2 $\pm$ 6.7 & 37.2 $\pm$ 8.7 & 36.7 $\pm$ 9.8 \\
        \quad CNN        & 35.4 $\pm$ 5.4 & 37.8 $\pm$ 6.9 & 38.7 $\pm$ 8.0 & 38.8 $\pm$ 8.8 \\
        \quad LSTM       & 36.4 $\pm$ 5.7 & 38.4 $\pm$ 7.5 & 38.0 $\pm$ 8.1 & 37.1 $\pm$ 8.4 \\
        \quad MtReNIRS-LP       & \textbf{39.1 $\pm$ 7.0} & \textbf{43.0 $\pm$ 9.7} & \textbf{44.2 $\pm$ 11.9} & \textbf{43.1 $\pm$ 12.1} \\
        \addlinespace
        \multicolumn{5}{l}{\textbf{SujinBak 2019} ($n=30$)} \\
        \quad Scratch-FT & 35.2 $\pm$ 3.0 & 38.0 $\pm$ 5.1 & 40.3 $\pm$ 6.4 & 41.9 $\pm$ 7.8 \\
        \quad Stat-SVM   & 32.6 $\pm$ 3.9 & 37.4 $\pm$ 7.1 & 39.5 $\pm$ 9.5 & 40.8 $\pm$ 10.6 \\
        \quad CNN        & 37.8 $\pm$ 6.9 & 42.7 $\pm$ 10.8 & 46.7 $\pm$ 12.6 & 48.9 $\pm$ 14.1 \\
        \quad LSTM       & 37.7 $\pm$ 6.6 & 41.1 $\pm$ 9.6 & 43.5 $\pm$ 11.0 & 45.2 $\pm$ 12.0 \\
        \quad MtReNIRS-LP       & \textbf{40.4 $\pm$ 8.1} & \textbf{46.2 $\pm$ 11.9} & \textbf{50.0 $\pm$ 14.2} & \textbf{53.0 $\pm$ 15.1} \\
        \bottomrule
    \end{tabular*}
\end{table}

\begin{figure*}[!t]
    \centering
    \includegraphics[width=0.90\textwidth]{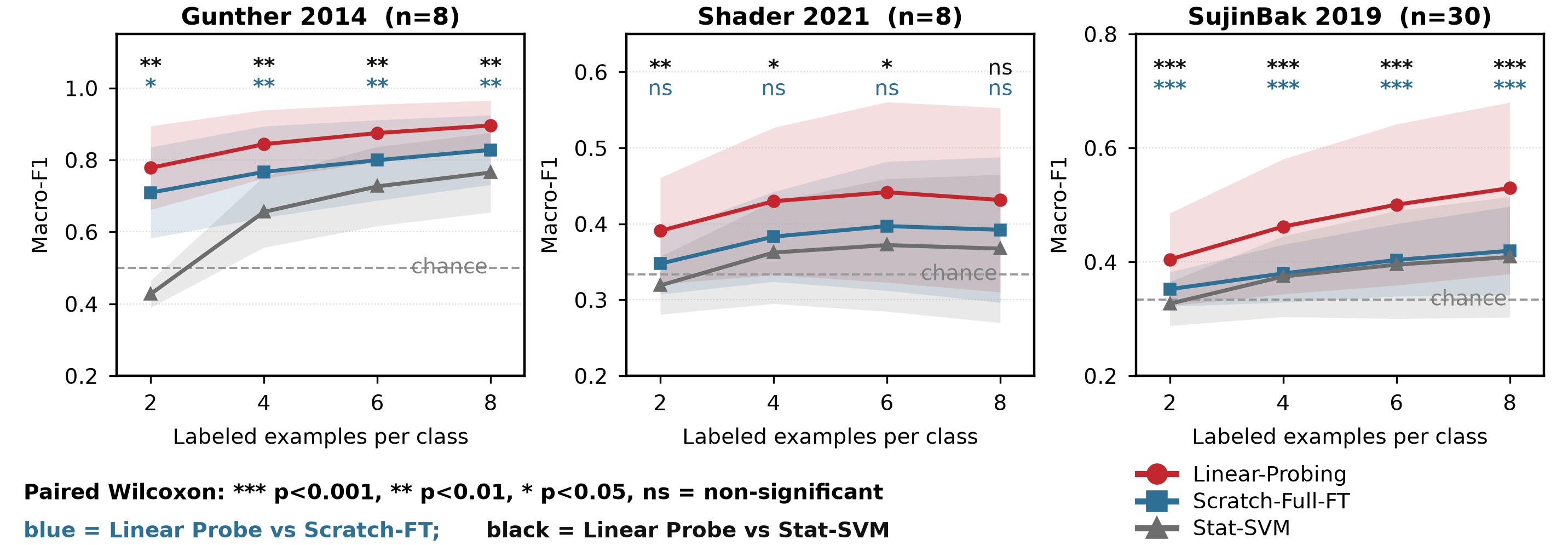}
    \caption{Within-subject few-shot macro-F1 performance across the three held-out evaluation datasets. The plots show the mean and standard deviation across subjects after averaging 50 random-seed repetitions within each subject; colored markers report paired two-sided Wilcoxon tests.}
    \label{fig:fewshot}
\end{figure*}

As shown in Table \ref{tab:benchmark2}, \acs{LP} had the highest point estimate among the tested methods in every dataset--shot condition. Relative to Scratch-FT, its paired gain ranged from
6.8 to 7.7 points on Gunther and was significant at all four shot counts
($p\leq0.0156$). The gain increased from 5.2 to 11.0 points on SujinBak and was
also significant throughout ($p\leq1.40\times10^{-5}$). Shader showed smaller
positive gains of 3.9--4.7 points that did not reach significance
($p=0.0547$--0.1484). \acs{LP} also exceeded \acs{Stat-SVM} significantly at every shot on
Gunther and SujinBak; on Shader, this comparison was significant at 2, 4, and
6 shots but not 8 shots. Figure~\ref{fig:fewshot} shows these subject-level
trends and paired tests. Overall, pretraining provided its clearest benefit
when a small labeled calibration set came from the target subject.

\begin{table}[!t]
    \caption{Few-shot accuracy comparison with prior work
    on the Gunther 2014 dataset. Prior work values in this table are reproduced from corresponding publications. For MtReNIRS, values are mean $\pm$ population standard deviation across subjects after averaging 50 random-seed repetitions within each subject; Highest values 
    are shown in bold.}
    \label{tab:gunther-prior-work}
    \centering
    \footnotesize
    \begin{tabular}{l@{\hspace{3pt}}c@{\hspace{4pt}}c@{\hspace{4pt}}c@{\hspace{4pt}}c}
        \toprule
        \textbf{Methods} & \multicolumn{4}{c}{\shortstack{\textbf{Gunther 2014} \\ \textbf{Mental arithmetic vs. rest}}} \\
        \midrule
        \textbf{Fully supervised} & \textbf{6-shot} & \textbf{12-shot} & \textbf{18-shot} & \textbf{24-shot} \\
        \midrule
        fNIRS-T~\cite{wang2022transformer} & 0.54 $\pm$ 0.15 & 0.58 $\pm$ 0.13 & 0.69 $\pm$ 0.10 & 0.88 $\pm$ 0.09 \\
        \specialrule{0.08em}{4pt}{2pt}
        \textbf{Pretrained models} & \textbf{2-shot} & \textbf{4-shot} & \textbf{6-shot} & \textbf{8-shot} \\
        \midrule
        MAE~\cite{jung2025fnirs, jung2025efrm} & 0.67 $\pm$ 0.26 & 0.67 $\pm$ 0.12 & 0.75 $\pm$ 0.10 & 0.75 $\pm$ 0.10 \\
        SSCL~\cite{jiang2021self} & 0.70 $\pm$ 0.12 & 0.73 $\pm$ 0.13 & 0.69 $\pm$ 0.22 & 0.69 $\pm$ 0.22 \\
        SleepFM~\cite{thapa2024sleepfm} & 0.75 $\pm$ 0.06 & 0.62 $\pm$ 0.09 & 0.60 $\pm$ 0.07 & 0.58 $\pm$ 0.13 \\
        \acs{LP} & \textbf{0.78 $\pm$ 0.11} & \textbf{0.85 $\pm$ 0.09} & \textbf{0.88 $\pm$ 0.08} & \textbf{0.90 $\pm$ 0.07} \\
        \bottomrule
    \end{tabular}
\end{table}

Table~\ref{tab:gunther-prior-work} places Gunther accuracy alongside
values reported by Jung et al.~\cite{jung2025fnirs}, benchmarking
MtReNIRS against a state-of-the-art fully supervised Transformer
classifier and existing fNIRS-only pretrained models. A $k$-shot
setting supplied exactly $k$ labeled examples per class.
\acs{LP} achieves the highest accuracy among the
fNIRS-only pretrained models across all matched 2--8-shot settings,
while also surpassing the fully supervised baseline at
6--24 shots. Since preprocessing, sampling, and evaluation details may differ across studies, these results provide contextual rather than controlled comparisons.

\section{Ablation Studies}

\subsection{Effect of Pretraining Duration}

\begin{figure}[!t]
    \centering
    \includegraphics[width=0.90\columnwidth]{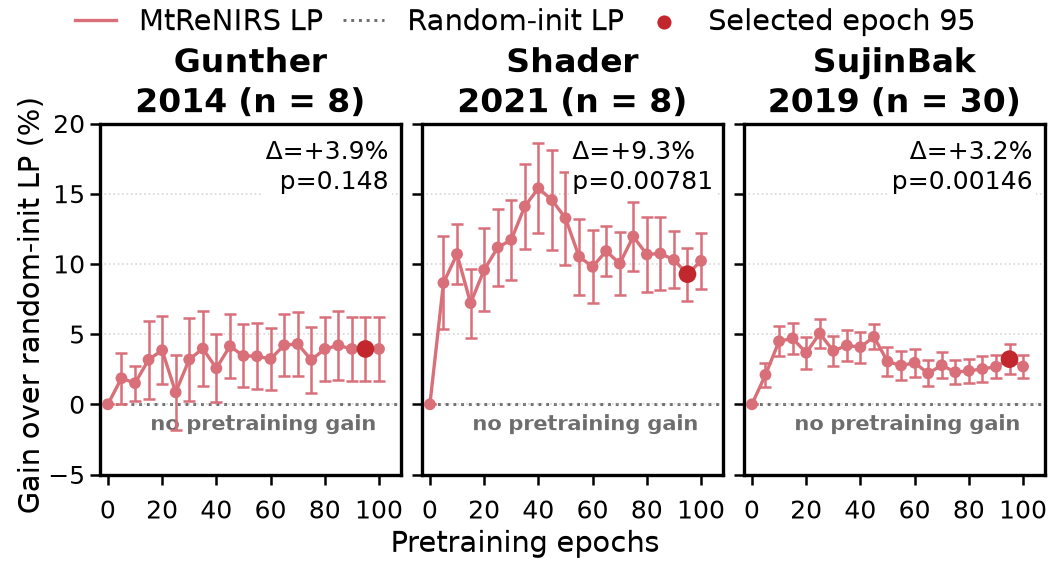}
    \caption{Macro-F1 gain of \acs{LP} over a randomly initialized frozen-encoder LP across pretraining epochs. Gains were computed within subjects, and error bars show the standard deviation across subjects. Legends report the gain and paired two-sided Wilcoxon $p$-value at the self-supervised checkpoint selected after epoch 95.}
    \label{fig:pretraining-gain}
\end{figure}

To measure the benefit of pretraining, we compared \acs{LP} with the same encoder architecture frozen at random initialization, with only the linear probe trained on the downstream task in both cases. We evaluated \acs{LP} at initialization and every five pretraining epochs. At the selected epoch 95 checkpoint, pretraining improved macro-F1 by 3.9 percentage points on Gunther, 9.3 on Shader, and 3.2 on SujinBak (Fig.~\ref{fig:pretraining-gain}). The gain was significant for Shader ($p=0.0078$) and SujinBak ($p=0.00146$), but not Gunther ($p=0.148$). Most of the gain emerged early, but additional pretraining did not improve gain monotonically across datasets. Gunther and SujinBak largely plateaued, whereas in Shader, downstream performance peaked around epoch 40 and declined thereafter, despite continued decreases in the pretraining reconstruction loss on both the training and validation sets.

\subsection{Effect of Temporal Context}
We compared token durations of 2, 5, and 10 seconds under 25\% masking during pretraining (see Section III-A for the rationale for the maximum duration). Each downstream example remained 10-s, and temporal-mean pooling kept the feature dimension fixed across token durations.

\begin{table}[!t] %
    \caption{Token-duration ablation under Benchmark 1 LOSO. Values are macro-F1 (\%) mean $\pm$ population standard deviation across held-out subjects. Boldface marks the highest mean within each dataset.}
    \label{tab:ablation-token-b1}
    \centering
    \scriptsize
    \setlength{\tabcolsep}{3pt}
    \begin{tabular*}{\columnwidth}{@{\extracolsep{\fill}}lccc@{}}
        \toprule
        Duration & \shortstack{Gunther \\ 2014} & \shortstack{Shader \\ 2021} & \shortstack{SujinBak \\ 2019} \\
        \midrule
        2 s  & $78.0 \pm 10.3$ & $36.6 \pm 6.6$ & $62.3 \pm 18.1$ \\
        5 s  & $82.0 \pm 8.7$  & $43.1 \pm 12.6$ & $65.0 \pm 19.3$ \\
        10 s & $\mathbf{87.3 \pm 7.8}$ & $\mathbf{44.5 \pm 9.4}$ & $\mathbf{67.9 \pm 17.7}$ \\
        \bottomrule
    \end{tabular*}
\end{table}

The 10-s model had the highest LOSO mean on all three datasets
(Table~\ref{tab:ablation-token-b1}). Within-subject Benchmark 2 showed the same trend, with the 10-s model achieving the highest mean across all three datasets and four shot counts (results not shown). This trend is consistent with prior fNIRS studies showing benefits from longer temporal context for capturing slow hemodynamic response \cite{wang2022transformer,jung2025efrm,jung2025fnirs}.

\subsection{Effect of Anatomical Correspondence}
Since a generalizable model should capture anatomical correspondence across different pretraining tasks, we tested whether the model learned this correspondence from the positional encodings seen during pretraining. We compared the proposed model (aligned) with a shuffled-coordinate baseline (shuffled), where spatial coordinates were randomly permuted among the selected chromophore channels within each crop. The permutation preserved every waveform, HbO/HbR identity, and coordinate value but reassigned each coordinate
to a different signal channel. Both
encoders received the correct held-out coordinates during downstream evaluation,
so this experiment isolated correspondence during pretraining.

\begin{table}[!t]
    \caption{Montage-aligned and coordinate-shuffled pretraining, Benchmark 2, \acs{LP}. Macro-F1 (\%) mean $\pm$ population standard deviation across subjects after averaging 50 repetitions within subject. Boldface marks highest mean in each paired comparison.}
    \label{tab:ablation-montage}
    \centering
    \scriptsize
    \setlength{\tabcolsep}{2pt}
    \begin{tabular*}{\columnwidth}{@{\extracolsep{\fill}}lcccc@{}}
        \toprule
        Montage & 2-shot & 4-shot & 6-shot & 8-shot \\
        \midrule
        \multicolumn{5}{l}{\textit{Gunther 2014}} \\
        Aligned  & $\mathbf{77.8 \pm 11.6}$ & $\mathbf{84.4 \pm 9.5}$ & $\mathbf{87.5 \pm 8.0}$ & $\mathbf{89.6 \pm 7.0}$ \\
        Shuffled & $67.6 \pm 11.9$ & $74.1 \pm 10.7$ & $78.6 \pm 8.3$ & $81.8 \pm 8.0$ \\
        \addlinespace[2pt]
        
        \multicolumn{5}{l}{\textit{Shader 2021}} \\
        Aligned  & $\mathbf{39.1 \pm 7.0}$ & $\mathbf{43.0 \pm 9.7}$ & $\mathbf{44.2 \pm 11.9}$ & $\mathbf{43.1 \pm 12.1}$ \\
        Shuffled & $37.3 \pm 4.1$ & $39.9 \pm 6.8$ & $41.2 \pm 9.3$ & $40.6 \pm 11.2$ \\
        \addlinespace[2pt]
        
        \multicolumn{5}{l}{\textit{SujinBak 2019}} \\
        Aligned  & $\mathbf{40.4 \pm 8.1}$ & $\mathbf{46.2 \pm 11.9}$ & $\mathbf{50.0 \pm 14.2}$ & $\mathbf{53.0 \pm 15.1}$ \\
        Shuffled & $35.4 \pm 3.9$ & $39.1 \pm 6.5$ & $42.1 \pm 8.6$ & $44.3 \pm 10.0$ \\
        \bottomrule
    \end{tabular*}
\end{table}

Within-subject Benchmark~2 favored montage alignment in all 12 dataset--shot
conditions: coordinate shuffling reduced mean few-shot macro-F1 at every shot
on every dataset (Table~\ref{tab:ablation-montage}). The reduction was
7.8--10.3 points on Gunther (all $p=0.0078$) and 5.0--8.7 points on SujinBak
(all $p\leq2.05\times10^{-7}$). Shader declined by 1.7--3.1 points in the same
direction, but no comparison reached $p<0.05$. Cross-subject Benchmark 1 showed the same trend; the results suggest that montage alignment improves transfer by enabling the model to learn shared representations that generalize across tasks, subjects, and devices.

\subsection{Robustness to Reduced Sensor Coverage}
Sparse-sensor robustness is desirable for practical wearable BCIs, where
fewer sensors improve portability and comfort while simplifying setup \cite{bhagubai2024towards}.
However, reducing sensor count can make decoding more sensitive to noise
and poor channel quality. We therefore evaluated \acs{LP} under reduced
sensor coverage on SujinBak. Its symmetric sensorimotor montage minimizes regional confounds when sensors are removed. The reference montage contained 18 S-D pairs arranged into nine bilateral region groups. For each sparse setting tested (2, 4, 8, 12, or 16 S-D pairs), we generated nine balanced subsets in which each group appeared equally often. Performance was averaged over subjects.

\begin{figure}[!t]
    \centering
    \includegraphics[width=0.90\columnwidth]{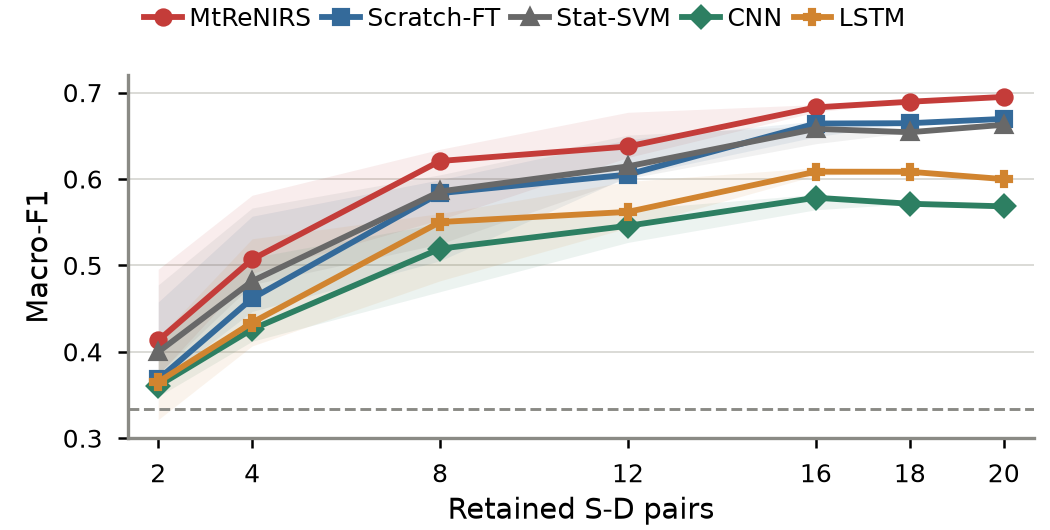}
    \caption{Reduced-montage robustness on SujinBak 2019 under LOSO. For 2--16 retained S-D pairs, lines show the median and shading the interquartile range across nine balanced bilateral subsets. The 18- and 20-pair points are single reference montages.}
    
    \label{fig:channel-reduction}
\end{figure}

As shown in Fig. \ref{fig:channel-reduction}, \acs{LP} had the highest median macro-F1 at every retained S-D pair
count. Its median increased from 0.413 with two pairs to 0.683 with 16 pairs,
compared with 0.689 for the symmetric 18-pair reference and 0.695 for the full
20-pair montage. At 16 pairs, \acs{Scratch-FT} and \acs{Stat-SVM} reached 0.664 and 0.658,
respectively. These results suggest that the pretrained representation remains robust under substantial reductions in sensor coverage, approaching full-montage performance with 16 retained S-D pairs. However, this analysis was performed on a single motor dataset and does not establish that the same robustness profile generalizes to other cortical regions or tasks.

\section{Discussion}

\textbf{Pretraining Benefits Personalized Adaptation More Than Zero-Calibration Transfer}: 
Across datasets, the largest and most consistent gains emerged in the within-subject few-shot setting rather than pure cross-subject transfer. This suggests that the pretrained encoder learns neurovascular features that generalize across tasks and acquisition settings, while subject-specific physiological variability remains a dominant source of error. Consequently, a small number of subject-specific labels appears sufficient to adapt a generic representation to a new user.

\textbf{Anatomical Alignment Is Important For Learning Transferable Representations}: 
The consistent degradation observed after coordinate shuffling suggests that transferability is derived not only from waveform structure but also from preserving anatomical correspondence across heterogeneous montages during pretraining.

\textbf{Limitations and Future Work}: Several limitations should be considered in this work. \textit{Temporal windowing limitation}: fNIRS measures a delayed and temporally extended hemodynamic response rather than neural activity directly, making representation quality dependent on the duration and timing of the analysis window. Although the 10-s configuration performed best in our experiments, different tasks and subjects may exhibit different response latencies, and future work should disentangle the effects of window duration from representation learning itself. \textit{Limited downstream datasets}: while pretraining was performed at relatively large scale for fNIRS, downstream evaluation was limited to the three held-out datasets used in prior work and may not fully characterize generalization across the broader range of tasks, acquisition settings, and experimental paradigms encountered in practice. \textit{Cross-device transfer not isolated}: Although pretraining successfully integrated recordings across multiple devices and montages, downstream evaluation could not isolate device effects because device, task, and montage differences co-occurred across datasets. Thus, the device invariance of the learned representations remains unclear. \textit{Cross-subject variability}: cross-subject transfer remains intrinsically challenging because anatomical variability, vascular dynamics, and sensor placement differ across individuals. Future work should evaluate larger and more diverse benchmarks, including controlled cross-device studies that vary acquisition hardware while holding other factors fixed. \textit{Multimodal pretraining}: Recent EEG--fNIRS work suggests that cross-modal pretraining can improve fNIRS-only downstream performance~\cite{jung2025efrm}. Future work could therefore test fusion strategies of complementary neurophysiological signals and explore the transferability and robustness of such multimodal frameworks.

\section{Conclusion}
We introduced MtReNIRS, a montage-aware foundation model for fNIRS pretrained across heterogeneous datasets spanning subjects, devices, tasks, and sensor layouts. The learned representations transferred to unseen datasets, with the largest benefits in few-shot subject adaptation. Together, the transfer and ablation results suggest that anatomically aligned self-supervision can serve as a scalable mechanism for learning reusable neurovascular representations, enabling future label-efficient fNIRS and multimodal BCI systems.

\section*{Code Availability}
Both MtReNIRS and the montage harmonization toolkit are publicly available: https://github.com/microsoft/MtReNIRS

\ifreferencesused

\fi

\end{document}